\documentclass[lettersize,journal]{IEEEtran}
\usepackage{amsmath,amsfonts,amsthm}
\usepackage{amssymb}
\usepackage[ruled,linesnumbered]{algorithm2e}
\usepackage{array}
\usepackage[caption=false,font=normalsize,labelfont=sf,textfont=sf]{subfig}
\usepackage{textcomp}
\usepackage{stfloats}
\usepackage{url}
\usepackage{verbatim}
\usepackage{graphicx}
\usepackage{cite}
\newtheorem{theorem}{Theorem}
\newtheorem{proposition}{Proposition}

\begin{document}
\title{Channel-Aware Vector Quantization for Robust Semantic Communication on Discrete Channels}

\author{Zian~Meng$^{\ast}$, Qiang~Li$^{\ast}$,~\IEEEmembership{Member,~IEEE}, Wenqian~Tang$^{\ast}$, Mingdie~Yan$^{\ast}$, and Xiaohu~Ge$^{\ast}$,~\IEEEmembership{Senior Member,~IEEE}\\
$^{\ast}$Huazhong University of Science and Technology, 430074 P. R. China
\thanks{Zian~Meng, Qiang~Li (Corresponding author), Wenqian~Tang, Mingdie~Yan, and Xiaohu~Ge are with Huazhong University of Science and Technology, 430074 P. R. China (emails: zian@hust.edu.cn, wqtang@hust.edu.cn, yanmingdie@hust.edu.cn, qli\_patrick@hust.edu.cn, xhge@mail.hust.edu.cn).}
\thanks{This work is partially supported by the Natural Science Foundation of China (NSFC) under grant U24A20212.}}



\maketitle

\begin{abstract}

Deep learning-based semantic communication has largely relied on analog or semi-digital transmission, which limits compatibility with modern digital communication infrastructures. Recent studies have employed vector quantization (VQ) to enable discrete semantic transmission, yet existing methods neglect channel state information during codebook optimization, leading to suboptimal robustness. To bridge this gap, we propose a channel-aware vector quantization (CAVQ) algorithm within a joint source-channel coding (JSCC) framework, termed VQJSCC, established on a discrete memoryless channel. In this framework, semantic features are discretized and directly mapped to modulation constellation symbols, while CAVQ integrates channel transition probabilities into the quantization process, aligning easily confused symbols with semantically similar codewords. A multi-codebook alignment mechanism is further introduced to handle mismatches between codebook order and modulation order by decomposing the transmission stream into multiple independently optimized subchannels. Experimental results demonstrate that VQJSCC effectively mitigates the digital cliff effect, achieves superior reconstruction quality across various modulation schemes, and outperforms state-of-the-art digital semantic communication baselines in both robustness and efficiency.

\end{abstract}


\section{Introduction}

\subsection{Background and Significance}
\IEEEPARstart{T}{he} evolution from 5G to 6G requires a paradigm shift beyond Shannon's communication framework, as the conventional focus on bits leads to spectral inefficiency and complex system architectures \cite{niu2022paradigm}. Conventional communication systems generally use separate source and channel coding (SSCC), following Shannon's separation theorem \cite{shannon1948mathematical}, which suggests it is optimal only if the block length is infinite. In real-world applications with finite block length, SSCC's effectiveness diminishes when channel errors exceed the correction capacity of the channel coding. As intelligent applications continue to expand, including the industrial Internet, vehicular communications, and augmented reality, SSCC becomes increasingly inadequate \cite{xiao2020toward}. Unlike syntax-focused systems that strictly separate source and channel coding, semantic communication utilizes shared knowledge bases and end-to-end semantic coding to minimize redundant data \cite{xie2021deep,getu2023tutorial} by focusing on the extraction and transmission of meaningful information rather than raw bits, thereby fostering more efficient and intelligent communication \cite{yang2022semantic}.

An imperative avenue in semantic communication research involves designing semantic coding schemes \cite{9771334}. Due to the challenges in applying strict algorithmic strategies to semantic extraction and processing, there has been a growing reliance on data-driven deep learning (DL) techniques for the advancement of semantic codecs \cite{farsad2018deep,xie2021deep,bourtsoulatze2019deep,PADC}. DL's nonlinear capacities enable the processing of semantic features from data \cite{lecun2015deep}. Essentially, semantics-driven codecs can be categorized as representation learning, as they focus on embedding meaningful features into low-dimensional forms while using joint source-channel coding (JSCC) to balance compactness and redundancy for error resilience \cite{tung2022deepjscc,huang2025d2jscc,choi2019neural}. By autonomously identifying the optimal redundancy level necessary to mitigate channel degradation, these methodologies advance into comprehensive end-to-end solutions that jointly tackle the challenges in semantic extraction, representation, and transmission.

Despite significant advancements in DL-based JSCC approaches, these methods predominantly assume continuous channel inputs with arbitrary precision \cite{weng2021speech, choi2019neural, bourtsoulatze2019image}. However, the fundamental requirements for transparent bit transmissions persist, with digital infrastructures remaining ubiquitously deployed in foreseeable technological evolution cycles \cite{evgenidis2023hybrid}. This leads to a critical conflict between analog JSCC assumptions and practical digital transmission mechanisms. In this context, ensuring seamless compatibility between semantic communication paradigms and existing digital channel constraints emerges as an unavoidable challenge. Recent studies have attempted to address this problem, yet many approaches fall short of strictly modeling fully discrete channels. For instance, some systems map semantic features to discrete constellations at the transmitter but still process noisy continuous constellation points at the receiver. In parallel, methods based on vector quantized variational autoencoder (VQ-VAE) enable fully discrete semantic representations but lack the capability for channel-aware end-to-end optimization, leading to degraded performance compared with analog baselines \cite{xie2020lite,tung2022deepjscc,he2023rate,nematiVQVAEEmpoweredWireless2023}. Motivated by this gap, this paper aims to resolve two pivotal questions:
\begin{enumerate}
\item \textit{How to design robust semantic communication architectures under discrete channel models?}
\item \textit{How to perform end-to-end optimization of digital semantic coding and modulation?}
\end{enumerate}

\IEEEpubidadjcol

\subsection{Related Works}
\subsubsection{DL-Enabled JSCC}
JSCC merges source and channel coding into a coherent process, achieving end-to-end optimality in communication systems. Initial JSCC methods \cite{farvardin1990study,nosratinia2003source,hamzaoui2005optimized,5563107} focused on specific source and channel distributions but faced implementation challenges due to high complexity. Recent advances in DL have revitalized JSCC through data-driven approaches, notably for wireless text transmission \cite{farsad2018deep} and image transmission \cite{bourtsoulatze2019deep}. DL-based JSCC exhibits two core attributes: source compression and end-to-end optimization. Source compression leverages deep neural networks (DNNs) to distill high-dimensional source data into lower-dimensional, critical features, thus facilitating the efficient use of bandwidth for transmission. Meanwhile, end-to-end optimization is achieved through gradient descent across simulated communication channels, which enables the automatic balancing of semantic fidelity with resilience to channel disruptions. These attributes collectively warrant the classification as JSCC.

Based on the two attributes mentioned above, the design of DL-based JSCC frameworks revolves around two fundamental aspects. The first aspect involves the DNN architecture, which demands customization to suit specific modalities and tasks. For instance, transformers are commonly utilized for transmitting text, whereas convolutional neural networks are used for image transmission. Additionally, various studies integrate rate adaptation techniques, such as dynamic masking \cite{PADC}, parameter lookup tables \cite{huang2025d2jscc}, or importance awareness mechanism \cite{liu2024ofdm}, to flexibly adjust the coding rate in response to channel conditions. The second aspect concerns the design of the simulated channel for training purposes. In most cases, the channel is modeled by either adding Gaussian noise to the features extracted by the DNN to simulate additive white Gaussian noise (AWGN) channels or by multiplying features with complex Gaussian random variables, followed by the addition of AWGN noise, to simulate Rayleigh fading channels \cite{xie2020deep,farsad2018deep,bourtsoulatze2019deep,weng2021semantic,dai2022nonlinear}. However, these schemes require full-resolution constellation transmission, creating a mismatch with existing digital communication infrastructures and hindering practical implementation \cite{wang2022constellation,xie2020lite,saidutta2021joint,dai2022nonlinear}. This challenge motivates the development of JSCC methods accommodating discrete representations for digital communications.

\subsubsection{Digital Semantic Coding and Modulation} 
To convert raw semantic data into compact discrete representations, scalar quantization techniques, which include uniform \cite{xie2020lite,huang2025d2jscc} and non-uniform types \cite{aibo2024digitalsc}, directly discretize continuous semantic features. Despite their computational efficiency, these techniques are inherently limited by the loss of information due to quantization noise, leading to degraded performance compared to analog semantic communication systems. Alternatively, stochastic coding frameworks \cite{boJointCodingmodulationDigital2024,tung2022deepjscc,bo2022learning} use variational inference to construct latent distributions, producing discrete symbols via stochastic sampling based on encoder-generated categorical distributions. Although capable of reaching information-theoretic bounds under specific channel conditions, these semi-digital methods still require continuous constellation inputs during decoding, which does not fully align with the discrete-alphabet requirements of standard digital decoders.

Inspired by the VQ-VAE model \cite{vqvae}, many emerging methods leverage vector quantization with codebook-based frameworks to implement fully digital semantic communication structures \cite{nematiVQVAEEmpoweredWireless2023,fuVectorQuantizedSemantic2023,xieRobustInformationBottleneck2023,hu2023robust}. In this approach, high-dimensional semantic data are encoded into continuous latent forms, then mapped onto discrete codebook entries based on the nearest distance criteria. This indexing aligns perfectly with digital communication pipelines by linking semantic symbols to modulation alphabets. However, the typical VQ-VAE framework does not consider channel states in its training objectives, which limits its adaptability to different channel conditions. Moreover, it links codebook size directly to modulation orders, restricting the flexibility in choosing modulation schemes while maintaining a constant quantization level.

\subsection{Contributions}
The main contributions of this paper are summarized as follows:
\begin{itemize}
    \item We propose VQJSCC, a robust digital semantic communication architecture that enables discrete signal processing at both the encoder and decoder. A DL-based image transmission system is established on a discrete memoryless channel (DMC), building upon the VQ-VAE backbone. Compared to existing analog and semi-digital JSCC schemes, VQJSCC provides channel-aware discrete semantic coding and accommodates arbitrary modulation schemes, addressing Question 1.
    \item To overcome the limitation of VQ methods that ignore channel state information (CSI), we propose an algorithm of channel-aware vector quantization (CAVQ).In view of the fact that the transmission error determines the upper bound of the reconstruction loss, CAVQ designs a loss function that explicitly incorporates the channel transition probability and aggregates the codewords with high error probabilities in the semantic space to minimize the transmission error, thereby achieving channel-aware joint optimization on discrete channel models. This addressed Question 2.
    \item When the codebook and modulation orders are mismatched, CAVQ fails to align the codebook with the channel, hindering systematic analysis. To address this, we propose a multi-codebook algorithm that divides the codebook index sequence into several independent discrete memoryless subchannels, each with its codebook that is individually optimized by CAVQ. This removes the necessity for a direct match between codebook size and modulation order, thus increasing the flexibility of digital semantic communication systems.
    \item Experimental results demonstrate that the proposed VQJSCC mitigates the cliff effect that is common in digital communication, surpassing state-of-the-art joint coding modulation methods in terms of peak-signal-to-noise ratio (PSNR) and computational efficiency. Our approach produces codebook distributions with a distinct pattern where modulation symbols that are easy to confuse are assigned with semantically similar codewords. Additionally, our method results in a more uniform codebook activation frequency, which ensures all codewords effectively contribute to information representation, thus achieving greater source entropy than vanilla VQ-VAE-based methodologies.

\end{itemize}

The remainder of this paper is organized as follows. Section \ref{sec:sysmodel} introduces the digital semantic transmission system and problem formulation. Section \ref{sec:methodology} presents the VQJSCC framework and its associated training process, with a focus on the CAVQ and multi-codebook strategies. Section \ref{sec:results} contains the analysis of experimental results, while Section \ref{sec:omega} presents the conclusions.

Notations: Scalars are denoted by lowercase and uppercase letters (e.g., $y$, $K$). Vectors are represented by boldface lowercase letters (e.g., $\mathbf{x}$). Fundamental sets include $\mathbb{R}$ (real numbers), $\mathbb{C}$ (complex numbers), and $\mathbb{Z}$ (integers), while customized sets are denoted by calligraphic letters (e.g., $\mathcal{X}$). The $\ell_2$-norm of a vector $\mathbf{x}$ is expressed as $\|\mathbf{x}\|_2$. The expectation of an expression with respect to the random variable $\mathbf{x}$ is denoted by $\mathbb{E}_{\mathbf{x}}[\cdot]$. The probability of a random event is represented by $P(\cdot)$. The Kronecker product of matrices $\mathbf{A}$ and $\mathbf{B}$ is denoted by $\mathbf{A} \otimes \mathbf{B}$.

\section{System Model and Problem Formulation} \label{sec:sysmodel}
\begin{figure*}[!t]
  \centering
  \includegraphics[width=0.75\textwidth]{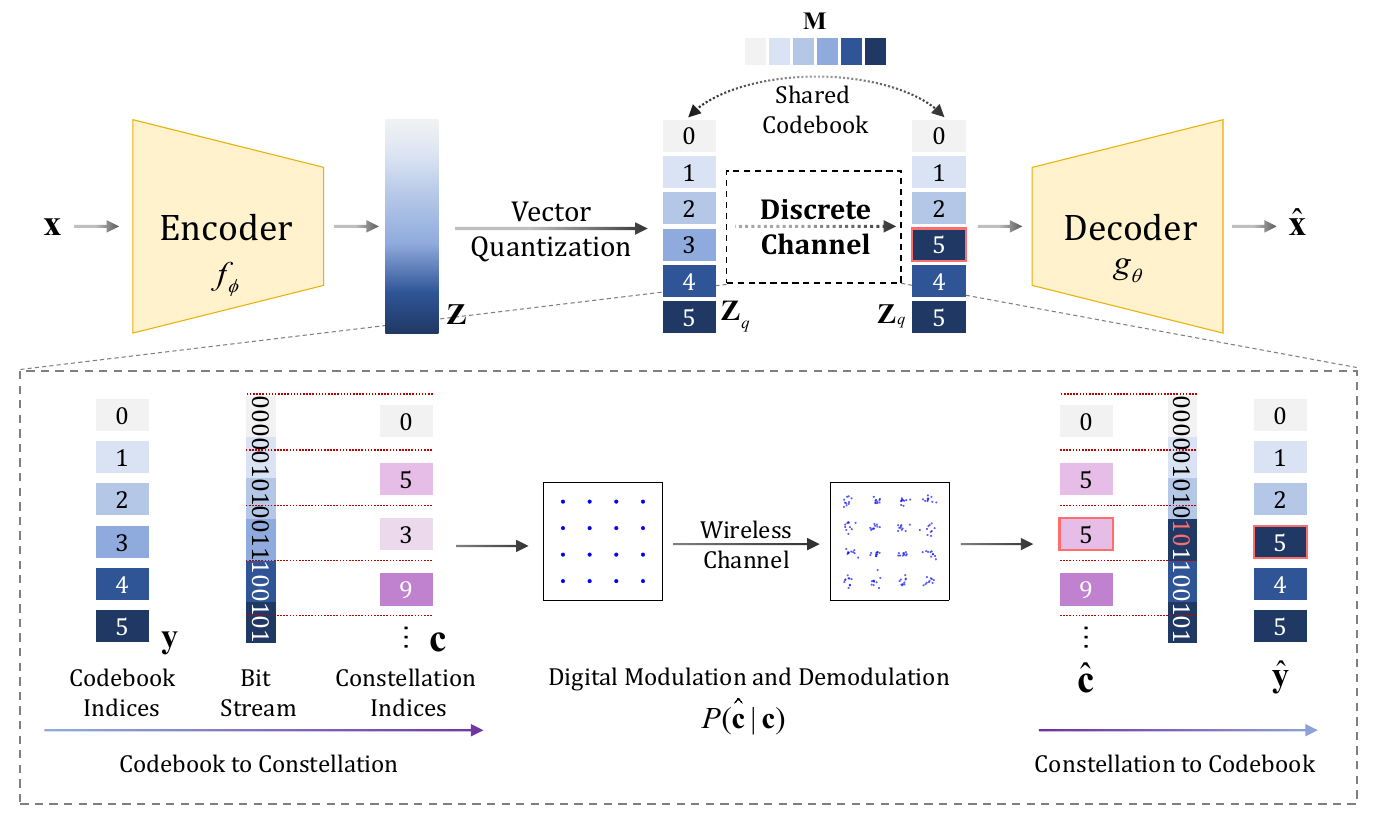}
  \caption{The proposed VQJSCC framework. The encoder \(f_{\phi}\) maps image \(\mathbf{x}\) to \(\mathbf{Z}\); vector quantization with a shared codebook \(\mathbf{M}\) yields \(\mathbf{Z}_q\) and indices \(\mathbf{y}\). The indices are bit-serialized and mapped to constellation indices \(\mathbf{c}\) (\(M=2^{m_c}\)), transmitted over the wireless channel and demodulated to \(\hat{\mathbf{c}}\), then pass through a discrete memoryless channel \(P(\hat{c}\mid c)\). The decoder \(g_{\theta}\) reconstructs \(\hat{\mathbf{x}}\) from \(\hat{\mathbf{y}}\).}

  \label{fig:vqvae}
\end{figure*}

Fig.~\ref{fig:vqvae} illustrates the proposed end-to-end digital semantic communication framework. The architecture is built upon the VQ-VAE backbone and consists of three main modules: semantic encoding/decoding, vector quantization, and digital modulation.

Let the input be a semantic source $\mathbf{x} \in \mathcal{X}$, where we focus on image transmission. The encoder $f_{\phi}: \mathcal{X} \to \mathbb{R}^{N \times d}$ extracts semantic features from the input image and outputs a matrix $\mathbf{Z} = [\mathbf{z}_1, \mathbf{z}_2, \ldots, \mathbf{z}_N]$ containing $N$ continuous $d$-dimensional vectors, where $\phi$ denotes the trainable encoder parameters. Here, $N$ corresponds to the number of transmission symbols, and $d$ represents the semantic embedding dimension per symbol. This process maps raw data into a compact semantic space $\mathbf{z}_i \in \mathbb{R}^d$.

To enable digital transmission, the system employs a trainable codebook $\mathbf{M} = [\mathbf{m}_1, \mathbf{m}_2, \ldots, \mathbf{m}_K] \in \mathbb{R}^{K \times d}$ for vector quantization. Each $\mathbf{z}_i$ is replaced by its nearest codeword $\mathbf{m}_{y_i}$ via Euclidean distance search:
\begin{align}
y_i = \arg\min_{j \in \{1,\ldots,K\}} \|\mathbf{z}_i - \mathbf{m}_j\|_2.
\end{align}
This yields quantized features $\mathbf{Z}_q = [\mathbf{m}_{y_1}, \ldots, \mathbf{m}_{y_N}]$ and a discrete index sequence $\mathbf{y} = [y_1, \ldots, y_N] \in \{1,\ldots,K\}^N$. The codebook size $K$ dictates both quantization granularity and transmission rate, since each index requires $\lceil\log_2 K\rceil$ bits. As the system follows the joint source-channel coding (JSCC) paradigm, the indices bypass channel coding and are directly passed to the modulation stage.

The digital modulation module maps $\mathbf{y}$ into constellation symbols. When the codebook size $K$ matches the modulation order $M$, the mapping is straightforward: each codebook index can be directly associated with a unique constellation point. However, in general $K \neq M$, a direct one-to-one mapping is infeasible. To ensure efficient use of the discrete representation, the index sequence $\mathbf{y}$ is first converted into a bitstream $\mathbf{b}=[b_0,\dots,b_{Nm_b}] \in \{0,1\}^{Nm_b}$, where $m_b = \lceil\log_2 K\rceil$. This bitstream is then partitioned into $L$ groups, each corresponding to a modulation symbol, yielding a constellation index sequence $\mathbf{c} = [c_1,\ldots,c_L] \in \{1,\ldots,M\}^L$, where $M = 2^{m_c}$ is the modulation order with $m_c$ bits per symbol:
\begin{align}
L = \left\lceil \frac{Nm_b}{m_c} \right\rceil.
\end{align}
Each index $c_i$ is then mapped to a constellation point $s_{c_i} \in \mathcal{S} \subset \mathbb{C}$, and the transmitted baseband signal is expressed as $\mathbf{s} = [s_{c_1}, \ldots, s_{c_L}] \in \mathbb{C}^L$.

The wireless channel introduces fading and additive noise, such that the received signal is modeled as
\begin{align}
\mathbf{r} = h\mathbf{s} + \mathbf{n},
\end{align}
where $h \in \mathbb{C}$ is the complex channel gain and $\mathbf{n} \sim \mathcal{CN}(0,\sigma^2\mathbf{I})$ is AWGN. With perfect channel state information (CSI) at the receiver, channel equalization can be applied to remove the effect of $h$, yielding an equivalent AWGN channel model.

Demodulation is performed using a detection function $\mathcal{D}: \mathbb{C}^L \to \{1,\ldots,M\}^L$ that maps received symbols to estimated constellation indices:
\begin{align}
\hat{\mathbf{c}} = \mathcal{D}(\mathbf{r}) = [\hat{c}_1, \ldots, \hat{c}_L].
\end{align}
A nearest-neighbor rule is applied, assigning each received point to the closest constellation symbol in $\mathcal{S}$.

To abstract modulation and demodulation into a tractable digital model, the channel is represented as a DMC. For each transmitted symbol $c_l$, the channel is fully characterized by transition probabilities:
\begin{align}
h_{ij} = P(\hat{c}_l=j \mid c_l=i), \quad \forall i,j \in \{1,\ldots,M\}.
\end{align}
These form the transition matrix $\mathbf{H} = [h_{ij}] \in [0,1]^{M \times M}$. Owing to the memoryless property, different symbols are independent:
\begin{align}
P(\hat{\mathbf{c}}|\mathbf{c}) = \prod_{l=1}^L P(\hat{c}_l|c_l).
\end{align}

Finally, the estimated bitstream $\hat{\mathbf{b}}$ is reconstructed from $\hat{\mathbf{c}}$, and mapped back to $\hat{\mathbf{y}} \in \{1,\ldots,K\}^N$. The shared codebook provides the quantized features $\hat{\mathbf{Z}}_q = [\mathbf{m}_{\hat{y}_1}, \ldots, \mathbf{m}_{\hat{y}_N}]$, which are decoded by $g_{\theta}: \mathbb{R}^{N \times d} \to \mathcal{X}$ to yield the final reconstruction $\hat{\mathbf{x}}$.

This digital semantic communication system is trained end-to-end as an image reconstruction task. The overall learning objective is to minimize the expected mean squared error (MSE) between the original image $\mathbf{x}$ and the reconstructed image $\hat{\mathbf{x}}$:
\begin{align}
\min_{\phi, \theta, \mathbf{M}} \; \mathbb{E}_{\mathbf{x}} \left[ \|\mathbf{x} - \hat{\mathbf{x}}\|_2^2 \right].
\end{align}
While the vanilla VQ-VAE framework employs the straight-through estimator (STE) to enable gradient propagation across the discrete quantization stage, it remains insufficient for digital semantic communication. In particular, it fails to capture and mitigate the impact of channel-induced symbol errors in the discrete latent space. Addressing this challenge requires channel-aware optimization, which is the focus of the next section.

\section{Channel-Aware Vector Quantization for Digital JSCC} \label{sec:methodology}

\subsection{Channel-Aware Codebook Optimization}
A key challenge in digital semantic communication is to align the latent semantic space with the statistical characteristics of the physical channel. While analog JSCC frameworks achieve this alignment implicitly through end-to-end training, digital systems require explicit mechanisms because vector quantization introduces discrete symbols that are directly vulnerable to channel impairments. This mismatch is referred to as the latent-channel distribution matching problem.

To illustrate, consider conventional digital modulation such as Quadrature Phase Shift Keying (QPSK). Noise in the channel makes neighboring constellation points more prone to confusion than distant ones. Gray coding is often applied so that adjacent points differ by only one bit, thereby reducing the bit error rate by aligning the modulation structure with channel error patterns. This principle highlights the importance of adapting discrete representations to channel statistics.

In digital semantic communication, a similar challenge arises. Channel noise may corrupt the transmitted codebook indices $\mathbf{y}$ into erroneous $\hat{\mathbf{y}}$, causing discrepancies between the reconstructed features $\hat{\mathbf{Z}}_q$ and the intended features $\mathbf{Z}_q$. These symbol errors propagate through the decoder, resulting in semantic distortion in the output $\hat{\mathbf{x}}$. To address this, it is essential to characterize how such errors influence the end-to-end reconstruction performance. For this purpose, we define the transmission error as
\begin{align}
\mathcal{L}_{t} = \mathbb{E}_{\mathbf{x},\hat{\mathbf{y}}}\left[ \|\mathbf{Z} - \hat{\mathbf{Z}}_q\|_2^2 \right].
\end{align}
The following theorem establishes that $\mathcal{L}_{t}$ provides an upper bound on the end-to-end reconstruction loss.

\begin{theorem}\label{the:upper}
The reconstruction loss is bounded as
\begin{align}
\mathbb{E}_{\mathbf{x,\hat{x}}}\left[ \|\mathbf{x} - \hat{\mathbf{x}}\|_2^2 \right] 
\leq 2\mathbb{E}_{\mathbf{x}}\left[ \|\mathbf{x} - g_\theta(\mathbf{Z})\|_2^2 \right] 
+ 2C_0^2 \mathcal{L}_{t},
\end{align}
where $C_0$ is the Lipschitz constant of the decoder $g_\theta$.
\end{theorem}

The proof is provided in Appendix~\ref{apx:t1}. Intuitively, the first term represents the intrinsic reconstruction error of the semantic encoder-decoder pair without channel noise, which becomes negligible after training. Thus, system performance is dominated by the transmission error $\mathcal{L}_t$, making its minimization the primary design goal.

By expanding $\mathcal{L}_t$, we obtain the following tractable formulation.

\begin{proposition}\label{prop:transloss}
The transmission error can be expressed as
\begin{align}
\mathcal{L}_t = \mathbb{E}_{\mathbf{x}} \left[ 
\sum_{n=1}^N \sum_{k=1}^{K} 
P(\hat{y}_n = k \mid y_n)\, 
\| \mathbf{z}_n - \mathbf{m}_{k} \|_2^2 \right],
\end{align}
where $N$ is the sequence length of discrete symbols $\mathbf{y}$, $K$ is the codebook size with $\mathbf{m}_k$ as the $k$-th codeword, and $P(\hat{y}_n = k \mid y_n)$ denotes the transition probability between indices.
\end{proposition}

The proof is given in Appendix~\ref{apx:p1}. Proposition~\ref{prop:transloss} shows that transmission error minimization can be directly integrated into gradient-based optimization. For the $i$-th codeword $\mathbf{m}_i \in \mathbf{M}$, the gradient update with respect to $\mathcal{L}_t$ is
\begin{align}
\nabla_{\mathbf{m}_{i}} \mathcal{L}_{t} 
= 2\mathbb{E}_{\mathbf{x}} \left[ 
\sum_{n=1}^N P(\hat{y}_n=i \mid y_n)\,(\mathbf{m}_{i}-\mathbf{z}_{n}) \right].
\end{align}
Geometrically, this update pulls $\mathbf{m}_i$ closer to features $\mathbf{z}_n$ that are frequently confused with index $i$ during transmission, thereby enforcing semantic similarity among error-prone codewords. This alignment improves robustness against channel-induced symbol errors.

\textbf{Remark 1}: The transmission error $\mathcal{L}_t$ generalizes the codebook loss used in the vanilla VQ-VAE. In the absence of channel errors, where the transition probability is:
\begin{align}
    P(\hat{y}_n = k \mid y_n) = \begin{cases}
    1, & \text{if } k = y_n, \\
    0, & \text{otherwise},
\end{cases}
\end{align}
$\mathcal{L}_t$ reduces to the vanilla VQ-VAE codebook loss:
\begin{align}
    \mathcal{L}_t = \mathbb{E}_{\mathbf{x}} \left[ \sum_{n=1}^N \| \mathbf{z}_n - \mathbf{m}_{y_n} \|_2^2 \right] = \mathbb{E}_{\mathbf{x}} \left[ \| \mathbf{Z} - \mathbf{Z}_q \|_2^2 \right] = \mathcal{L}_c.
\end{align}
This demonstrates that the vanilla VQ-VAE codebook loss is a special case of the proposed transmission error, applicable only when the channel is error-free. In contrast, our proposed method accounts for channel-induced errors, offering greater generality and enabling channel-aware codebook optimization. This approach effectively mitigates the impact of channel noise, enhancing the performance and reliability of digital semantic communication systems.

\subsection{Multi-Codebook and Sub-Channel}

\begin{figure*}[t!]
  \centering
  \includegraphics[width=\textwidth]{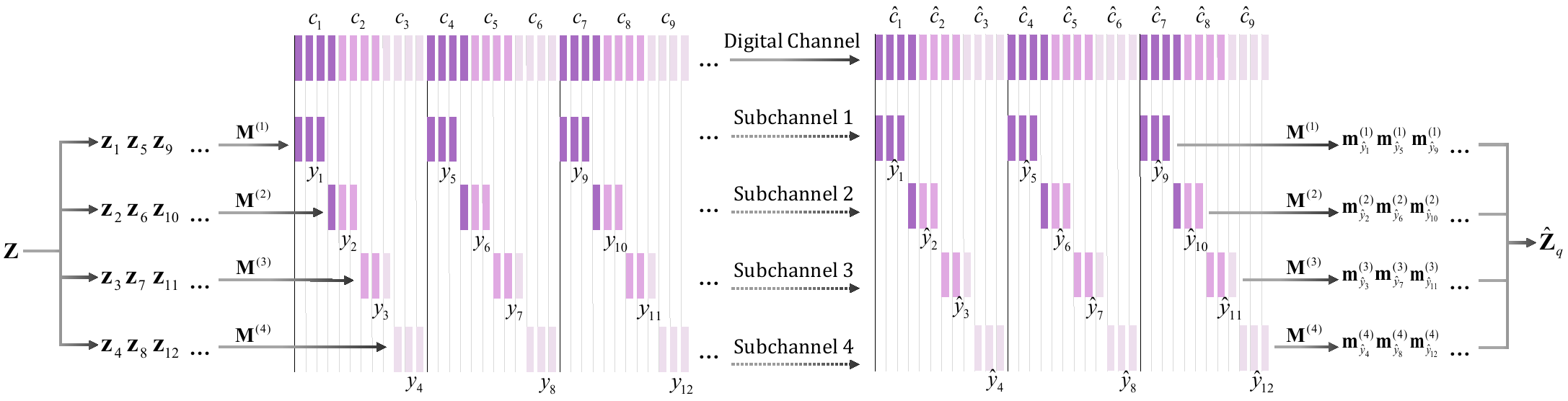}
  \caption{Subsequence decomposition and multi-codebook mapping address mismatched codebook and modulation orders. The index stream is divided based on the repeating bit-alignment period into separate subsequences. Each subsequence extracts a consistent bit slice pattern from sequential constellation symbols, creating an i.i.d. discrete memoryless subchannel via the digital channel.}
  \label{fig:subchannel}
\end{figure*}

When the codebook order $m_b$ (with codebook size $K = 2^{m_b}$) does not match the modulation order $m_c$, the binary representation of codebook indices must be reorganized to fit into modulation symbols. For instance, if each codebook index is represented by 6 bits ($m_b=6$), but the modulation uses 4 bits per symbol ($m_c=4$), some indices will be split across symbols. This reorganization creates statistical dependencies between adjacent codebook indices, breaking the memoryless property that underpins our analysis. As a result, the transmission error cannot be computed directly.

As shown in Fig. \ref{fig:subchannel}, to restore tractability, we introduce a sequence decomposition method that divides the original sequence into multiple independent subsequences. Each subsequence can then be modeled as a discrete memoryless subchannel with its own transition probability matrix.

\paragraph{Step 1: Constructing subsequences.}
We first define the grouping period as
\begin{align}
T = \mathrm{lcm}(m_b,m_c),
\end{align}
the least common multiple of the codebook and modulation orders. Intuitively, $T$ is the smallest number of bits after which the alignment pattern between codebook indices and modulation symbols repeats.

The original index sequence $\mathbf{y}=(y_1,\ldots,y_N)$ is divided into frames of length $T/m_b$ indices. Since each frame contains exactly $T$ bits, the bit alignment across frames is identical. We then extract $N_s = T/m_b$ subsequences by collecting the $i$-th index of every frame:
\begin{align}
\mathbf{y}^{(i)} = \bigl(y_{i+kT/m_b}\bigr)_{k=0}^{N m_b/T - 1}, 
\quad i=1,\ldots,N_s.
\end{align}

Each subsequence $\mathbf{y}^{(i)}$ satisfies two key properties:
\begin{itemize}
    \item \textbf{Independence:} Different frames are statistically independent because the underlying physical channel is a DMC.
    \item \textbf{Identical distribution:} Codebook indices occupying the same temporal slot in different frames share the same bit composition pattern, due to the periodicity introduced by $T$.
\end{itemize}
Together, these properties imply that each subsequence $\mathbf{y}^{(i)}$ can be regarded as an independent discrete memoryless subchannel.

\paragraph{Step 2: Transition matrices of subchannels.}
Within a subsequence, each element $y^{(i)}$ may draw its bits from multiple modulation symbols. Let $\tau_i$ denote the number of distinct bit-segments that contribute to $y^{(i)}$, and let $\mathbf{b}^{(i)}_t$ be the $t$-th such segment. For example, if $m_b=3$ and $m_c=4$, one subsequence element may gather all 3 bits from a single modulation symbol ($\tau_i=1$), while another may gather its bits from two different symbols ($\tau_i=2$).

For each bit-segment, we define a marginal transition matrix $\mathbf{G}^{(i)}_t$, whose $(p,q)$ entry represents the probability that an input segment with value $p-1$ is received as $q-1$. To compute this, we introduce the masked constellation set
\begin{align}
\mathcal{C}(\mathbf{b}^{(i)}_t) = 
\bigl\{ c \in \{1,\ldots,2^{m_c}\} \,\big|\, \mathrm{mask}^{(i)}_t(c)=\mathbf{b}^{(i)}_t \bigr\},
\end{align}
which contains all modulation indices consistent with the segment $\mathbf{b}^{(i)}_t$. Using this, the marginal matrix entries are
\begin{align}
\mathbf{G}^{(i)}_t(p,q) =
\frac{\sum\limits_{\hat{c} \in \mathcal{C}(\hat{\mathbf{b}}^{(i)}_t)}
      \sum\limits_{c \in \mathcal{C}(\mathbf{b}^{(i)}_t)}
      P(\hat{c}|c)P(c)}
     {\sum\limits_{c \in \mathcal{C}(\mathbf{b}^{(i)}_t)} P(c)}.
     \label{eq:margin}
\end{align}
Here, $P(\hat{c}|c)$ is the physical channel’s symbol transition probability, and $P(c)$ is the distribution of transmitted constellation indices.

Because the $\tau_i$ segments of $y^{(i)}$ originate from distinct modulation symbols, their errors are independent. Therefore, the composite subchannel transition matrix is the Kronecker product of its marginal matrices:
\begin{align}
\mathbf{H}^{(i)} = \bigotimes_{t=1}^{\tau_i} \mathbf{G}^{(i)}_t.
\end{align}

\paragraph{Worked Example.}
Consider $m_b=3$ and $m_c=4$. Each codebook index has 3 bits, while each modulation symbol carries 4 bits.  
\begin{itemize}
    \item For the first subsequence, $y_1$ uses bits $b_1b_2b_3$ from $c_1$. Since all bits come from a single modulation symbol, $\tau_1=1$ and $\mathbf{H}^{(1)}=\mathbf{G}^{(1)}_1$.  
    \item For the second subsequence, $y_2$ uses $b_4$ (the last bit of $c_1$) together with $b_5b_6$ (the first two bits of $c_2$). In this case, $\tau_2=2$, so the transition matrix is $\mathbf{H}^{(2)}=\mathbf{G}^{(2)}_1 \otimes \mathbf{G}^{(2)}_2$.  
\end{itemize}

By repeating this procedure for all subsequences, the original physical channel is decomposed into a set of independent subchannels $\{\mathbf{H}^{(i)}\}_{i=1}^{N_s}$. These subchannels can then be individually optimized using separate codebooks, greatly enhancing flexibility when $m_b$ and $m_c$ are mismatched.


\begin{algorithm}[htbp]
\caption{Computing Independent Subchannel Transition Probability Matrices}
\label{alg:remap}
\KwIn{Codebook order $m_b$, modulation order $m_c$, modulation symbol transition probabilities $P(\hat{c}\mid c)$, and modulation symbol distribution $P(c)$.}
\KwOut{Independent subchannel transition probability matrices $\{\mathbf{H}^{(i)}\}_{i=1}^{N_s}$.}

Compute the grouping period: $T \gets \mathrm{lcm}(m_b, m_c)$\;
Determine the number of subsequences: $N_s \gets T/m_b$\;

\For{$i \in \{1,2,\ldots,N_s\}$}{
    Identify $\tau_i$ bit-segments of constellation indices forming the $i$-th subsequence\;
    \For{$t \in \{1,2,\ldots,\tau_i\}$}{
        Construct the marginal transition probability matrix $\mathbf{G}_t^{(i)}$\;
        \For{$p,q \in \{1,\ldots,2^{\mathrm{len}(\mathbf{b}^{(i)}_t)}\}$}{
            Compute $\mathbf{G}_t^{(i)}(p,q)$ using (\ref{eq:margin})\;
        }
    }
    Form the composite subchannel matrix: $\mathbf{H}^{(i)} \gets \bigotimes_{t=1}^{\tau_i} \mathbf{G}_t^{(i)}$\;
}
\Return $\{\mathbf{H}^{(i)}\}_{i=1}^{N_s}$\;
\end{algorithm}

\subsection{Model Architecture and Training Method}
Leveraging CAVQ and the multi-codebook scheme, we design a robust image transmission model that can jointly optimize coding according to channel conditions. The trainable parameters include the encoder $f_\phi$ and decoder $g_\theta$ with parameters $\phi$ and $\theta$, as well as multiple codebooks. The model architecture and training algorithm are described as follows.

An important objective in our design is to make the coding rate controllable. In conventional VQ-VAE, each latent feature vector is quantized into a single codeword, leading to a fixed number of transmitted indices regardless of channel conditions. This rigidity is suboptimal in digital semantic communication, where the available channel capacity may vary. To provide flexibility, we introduce a hierarchical feature decomposition mechanism.

Specifically, given an input image $\mathbf{x}\in \mathbb{R}^{H\times W\times C}$, the encoder $f_\phi$ produces a latent semantic feature matrix
\begin{align}
\mathbf{Z} = [\mathbf{z}_1,\ldots,\mathbf{z}_N] \in \mathbb{R}^{N\times (d\cdot l)},
\end{align}
where $N$ is the number of feature vectors, $d$ is the base dimension (equal to the codeword dimension), and $l$ is the index depth. When $l=1$, each feature vector $\mathbf{z}_i$ is directly quantized into a single index. When $l>1$, $\mathbf{z}_i \in \mathbb{R}^{d\cdot l}$ is partitioned into $l$ sub-vectors of dimension $d$, all quantized by the same codebook. As a result, each spatial position yields $l$ discrete indices, and the total number of indices per image becomes $N\cdot l$. With a codebook order of $m_b$, corresponding to $m_b$ bits per index, the overall transmitted bit length equals $N\cdot l\cdot m_b$.

This hierarchical decomposition provides an effective mechanism to scale the coding rate: increasing $l$ produces longer index sequences, enabling richer semantic representations; decreasing $l$ reduces the number of transmitted indices, allowing operation under tighter channel capacity constraints. Thus, the same image can be encoded at different semantic rates, making the system adaptive to diverse channel conditions.

The feature matrix $\mathbf{Z}$ is further divided into $N_s$ subsequences $\{\mathbf{Z}^{(i)}\}_{i=1}^{N_s}$, each with $N_c=N/N_s$ feature vectors:
\begin{align}
\mathbf{Z}^{(i)}=[\mathbf{z}^{(i)}_1,\ldots,\mathbf{z}^{(i)}_{N_c}],
\end{align}
where zero-padding is applied if $N$ is not divisible by $N_s$. Each subsequence is quantized by a dedicated codebook $\mathbf{M}^{(i)}=[\mathbf{m}_1^{(i)},\ldots,\mathbf{m}_K^{(i)}]\in\mathbb{R}^{K\times d}$, where $K=2^{m_b}$ is determined by the codebook order $m_b$. Vector quantization is performed via nearest-neighbor search:
\begin{align}
y_j^{(i)} = \arg\min_{k}\|\mathbf{z}_j^{(i)}-\mathbf{m}_k^{(i)}\|_2,
\end{align}
yielding index sequences $\mathbf{y}^{(i)}\in\{1,\ldots,K\}^{N_c}$ and quantized vectors
\begin{align}
\mathbf{Z}_q^{(i)}=[\mathbf{m}^{(i)}_{y_1^{(i)}},\ldots,\mathbf{m}^{(i)}_{y_{N_c}^{(i)}}].
\end{align}

The indices are serialized into a bitstream $\mathbf{b}\in\{0,1\}^{Nm_b}$, then grouped according to the modulation order $m_c$ to form the sequence of modulation indices $\mathbf{c}=[c_1,\ldots,c_L]$, where $L=Nm_b/m_c$. Each $c_i$ is mapped to a constellation point for transmission. At the receiver, the demodulated sequence $\hat{\mathbf{c}}$ is modeled as passing through a DMC with transition matrix $\mathbf{H}\in\mathbb{R}^{2^{m_c}\times 2^{m_c}}$. Following Algorithm~\ref{alg:remap}, $N_s$ subsequences $\{\hat{\mathbf{y}}^{(i)}\}_{i=1}^{N_s}$ are reconstructed, and each is mapped back to quantized features via its codebook:
\begin{align}
\hat{\mathbf{Z}}_q^{(i)}=[\mathbf{m}^{(i)}_{\hat{y}_1^{(i)}},\ldots,\mathbf{m}^{(i)}_{\hat{y}_{N_c}^{(i)}}].
\end{align}
All subsequences are concatenated to form $\hat{\mathbf{Z}}_q\in\mathbb{R}^{N\times d}$, which is decoded by $g_\theta$ to reconstruct the image $\hat{\mathbf{x}}$.

The loss function replaces the vanilla VQ-VAE codebook loss with the analytical channel-aware transmission error:
\begin{align}
\mathcal{L}_{\text{ca}} = \sum_{i=1}^{N_s}\sum_{k=1}^{N_c}p_k^{(i)}
\sum_{j=1}^Kh_{y_k^{(i)}j}^{(i)}\,
\|\text{sg}[\mathbf{z}_k^{(i)}]-\mathbf{m}_j^{(i)}\|_2^2,
\end{align}
where $h^{(i)}_{y_k^{(i)}j}$ is the $(y_k^{(i)},j)$ element of subchannel matrix $\mathbf{H}^{(i)}$, $p_k^{(i)}$ is the occurrence probability of $y_k^{(i)}$, and $\text{sg}[\cdot]$ denotes the stop-gradient operator. The straight-through estimator (STE)~\cite{STE} is applied to enable gradient propagation through quantization.

To mitigate codebook collapse, we incorporate Clustering VQ-VAE (CVQ)~\cite{cvqvae}, which uses an exponential moving average (EMA) of codeword activations to re-anchor underutilized codewords. For the $k$-th codeword in codebook $\mathbf{M}^{(i)}$, the usage counter is updated as
\begin{align}
N_{k,t}^{(i)} = \gamma N_{k,t-1}^{(i)} + \frac{n_{k,t}^{(i)}}{N_c}(1-\gamma),
\end{align}
where $n_{k,t}^{(i)}$ counts the number of features assigned to $\mathbf{m}_k^{(i)}$. The anchor update is
\begin{align}
\alpha_{k,t}^{(i)} &= \exp\!\Big(-N_{k,t}^{(i)}K\frac{10}{1-\gamma}-\epsilon\Big),\\
\mathbf{m}_{k,t}^{(i)} &= (1-\alpha_{k,t}^{(i)})\mathbf{m}_{k,t-1}^{(i)}+\alpha_{k,t}^{(i)}\hat{\mathbf{z}}_{k,t}^{(i)},
\end{align}
where $\hat{\mathbf{z}}_{k,t}^{(i)}$ is the nearest feature to $\mathbf{m}_k^{(i)}$. This prevents codeword deactivation by shifting unused codewords toward dense feature regions. The complete training procedure is summarized in Algorithm~\ref{alg:vqjscc}.

\begin{algorithm}[t!]
\caption{Training VQJSCC Model}
\label{alg:vqjscc}
\KwIn{Dataset $\mathcal{D}$, codebook order $m_b$, modulation order $m_c$, channel transition matrix $\mathbf{H}$}
\KwOut{Optimized encoder parameters $\phi^*$, decoder parameters $\theta^*$, and codebooks $\{\mathbf{M}^{(i)}\}^*$}

\For{epoch $=1$ \KwTo MaxEpochs}{
    Sample a training image: $\mathbf{x} \sim \mathcal{D}$\;
    Encode semantic features: $\mathbf{Z} \gets f_\phi(\mathbf{x})$\;
    Partition into subsequences: $\{\mathbf{Z}^{(i)}\}_{i=1}^{N_s} \gets \text{Split}(\mathbf{Z},N_s)$\;
    Construct subchannel matrices: $\{\mathbf{H}^{(i)}\}_{i=1}^{N_s} \gets \text{Algorithm~\ref{alg:remap}}(\mathbf{H},m_b,m_c)$\;

    \For{$i=1$ \KwTo $N_s$}{
        Quantize features via codebook $\mathbf{M}^{(i)}$ to obtain index sequence $\mathbf{y}^{(i)}$\;
        Count codeword activations: $n_k^{(i)} \gets \sum_j \mathbb{I}(y_j^{(i)}=k)$ for $k=1,\ldots,K$\;
        Pass through subchannel: $\hat{\mathbf{y}}^{(i)} \sim P(\cdot|\mathbf{y}^{(i)},\mathbf{H}^{(i)})$\;
        Retrieve quantized features: $\hat{\mathbf{Z}}_q^{(i)} \gets \mathbf{M}^{(i)}[\hat{\mathbf{y}}^{(i)}]$\;
    }

    Concatenate subsequences: $\hat{\mathbf{Z}}_q \gets \text{Concat}(\{\hat{\mathbf{Z}}_q^{(i)}\}_{i=1}^{N_s})$\;
    Decode reconstruction: $\hat{\mathbf{x}} \gets g_\theta(\hat{\mathbf{Z}}_q)$\;

    Compute reconstruction loss: $\mathcal{L}_r \gets \|\mathbf{x}-\hat{\mathbf{x}}\|_2^2$\;
    Compute commitment loss: $\mathcal{L}_m \gets \sum_{i,j}\|\mathbf{z}_j^{(i)}-\text{sg}[\mathbf{m}^{(i)}_{y_j^{(i)}}]\|_2^2$\;
    Compute channel-aware loss: $\mathcal{L}_{\text{ca}} \gets \sum_{i,j,k}h_{y_j^{(i)}k}^{(i)}\|\text{sg}[\mathbf{z}_j^{(i)}]-\mathbf{m}_k^{(i)}\|_2^2$\;

    Update encoder: $\phi \gets \phi-\eta\nabla_\phi(\mathcal{L}_r+\beta\mathcal{L}_m)$\;
    Update decoder: $\theta \gets \theta-\eta\nabla_\theta\mathcal{L}_r$\;

    \For{$i=1$ \KwTo $N_s$}{
        \For{$k=1$ \KwTo $K$}{
            Update codeword usage: $N_k^{(i)} \gets \gamma N_k^{(i)}+(1-\gamma)\tfrac{n_k^{(i)}}{N_c}$\;
            Compute smoothing factor: $\alpha_k^{(i)} \gets \exp\!\big(-N_k^{(i)}K\tfrac{10}{1-\gamma}-\epsilon\big)$\;
            Find nearest feature: $\mathbf{z}_{\text{nearest}} \gets \arg\min_{\mathbf{z}\in\mathbf{Z}^{(i)}}\|\mathbf{z}-\mathbf{m}_k^{(i)}\|_2$\;
            Update codeword via CVQ: $\mathbf{m}_k^{(i)} \gets (1-\alpha_k^{(i)})\mathbf{m}_k^{(i)}+\alpha_k^{(i)}\mathbf{z}_{\text{nearest}}$\;
        }
        Gradient update of codebook: $\mathbf{M}^{(i)} \gets \mathbf{M}^{(i)}-\eta\nabla_{\mathbf{M}^{(i)}}\mathcal{L}_{\text{ca}}$\;
    }
}
\Return $\phi^*,\theta^*,\{\mathbf{M}^{(i)}\}^*$
\end{algorithm}

\section{Experimental Results} \label{sec:results}
\subsection{Experimental Settings}
\subsubsection{Dataset}
The CIFAR-10 dataset\cite{krizhevsky2009learning}, comprising 60,000 32x32 color images across 10 categories, is used for training and testing. The dataset is split into 50,000 training images and 10,000 test images.

\subsubsection{Parameter Settings}
Training is conducted over 200 epochs with a batch size of 128. The initial learning rate is set to $10^{-3}$, with a decay mechanism triggered if no improvement in validation performance is observed for 20 consecutive epochs. The signal-to-noise ratio (SNR) is dynamically sampled from the range [0, 18] dB for each training batch to simulate varying channel conditions. 

\subsubsection{Metric and Benchmark Methods}
To ensure fairness in comparing analog and digital communication paradigms, a unified criterion is adopted: evaluating image reconstruction quality at the same symbol rate. For digital communication, the symbol rate refers to the transmission rate of constellation points, where each symbol corresponds to a constellation point index. For analog communication, the symbol rate refers to the transmission rate of individual floating-point numbers in the feature vector. This comparison method is widely used in semantic communication research and is scientifically justified and reproducible \cite{PADC, xie2020lite, bourtsoulatze2019deep}.
\begin{itemize}
    \item \textbf{Analog:} Gaussian-distributed noise is added directly to the continuous features output by the encoder, corresponding to an analog AWGN channel.
    \item \textbf{Vanilla VQ-VAE with clustering:} The baseline VQ-VAE algorithm without channel awareness. The clustering mechanism \cite{cvqvae} is also used to ensure fair comparison.
    \item \textbf{JCM:} A state-of-the-art semi-digital JSCC joint coding-modulation (JCM) architecture proposed in \cite{boJointCodingmodulationDigital2024}.
    \item \textbf{Turbo-BPG:} The conventional SSCC communication using BPG as source coding and Turbo as channel coding.
\end{itemize}

\subsection{Analysis of Codebook Space}

\begin{figure}[t!]
  \centering
  \includegraphics[width=0.48\textwidth]{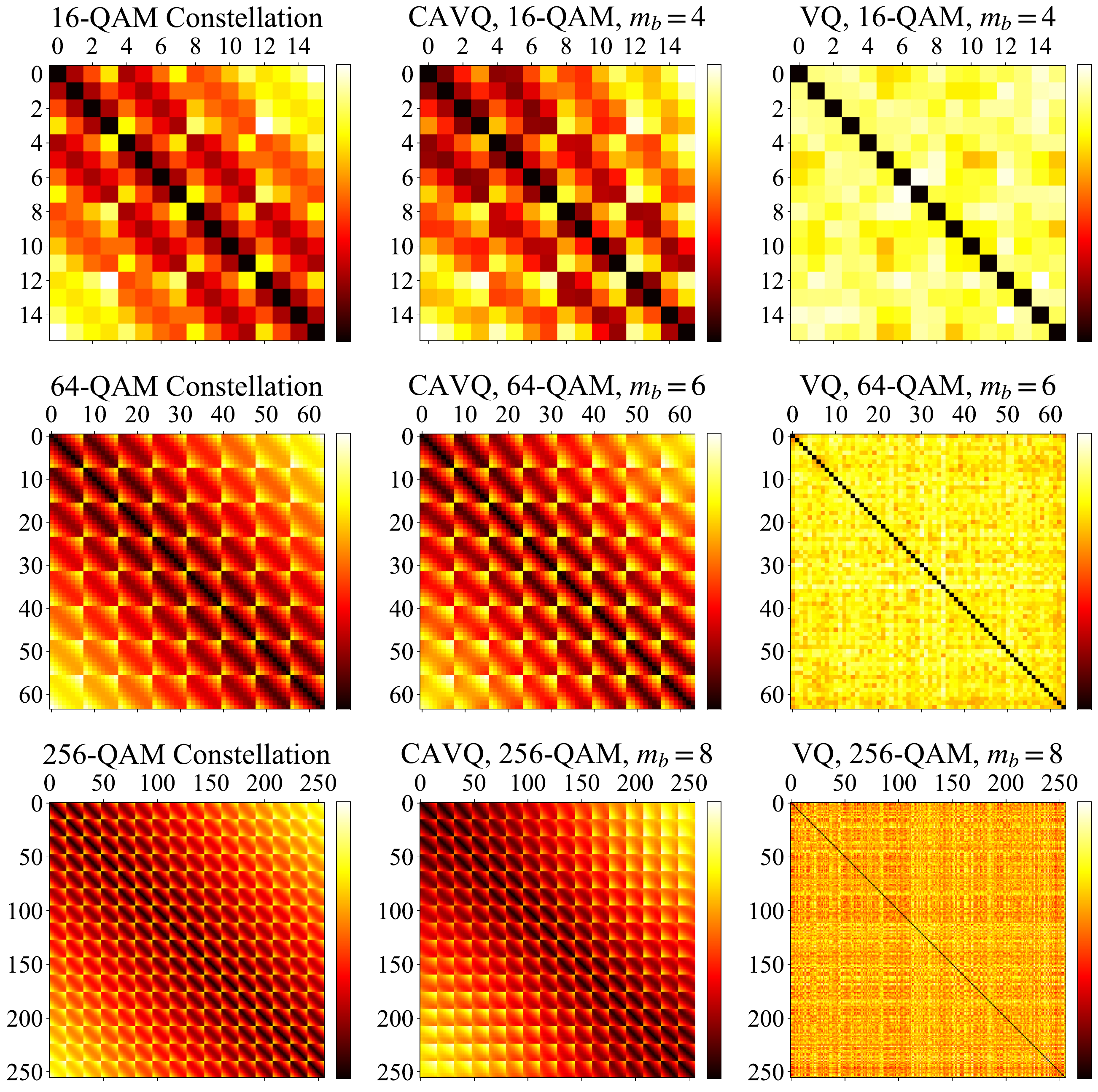}
  \caption{Heatmaps of pairwise distance matrices for modulation constellations (left), codebooks learned by the proposed CAVQ (middle), and codebooks from vanilla VQ-VAE (right) under 16-QAM, 64-QAM, and 256-QAM. Warmer colors indicate larger distances. CAVQ produces structured, constellation-like patterns aligned with modulation geometry, while vanilla VQ-VAE exhibits irregular distributions without channel awareness.}
  \label{fig:codebook_hotmap}
\end{figure}

\begin{figure}[t!]
  \centering
  \begin{minipage}{0.33\textwidth}
    \includegraphics[width=\linewidth]{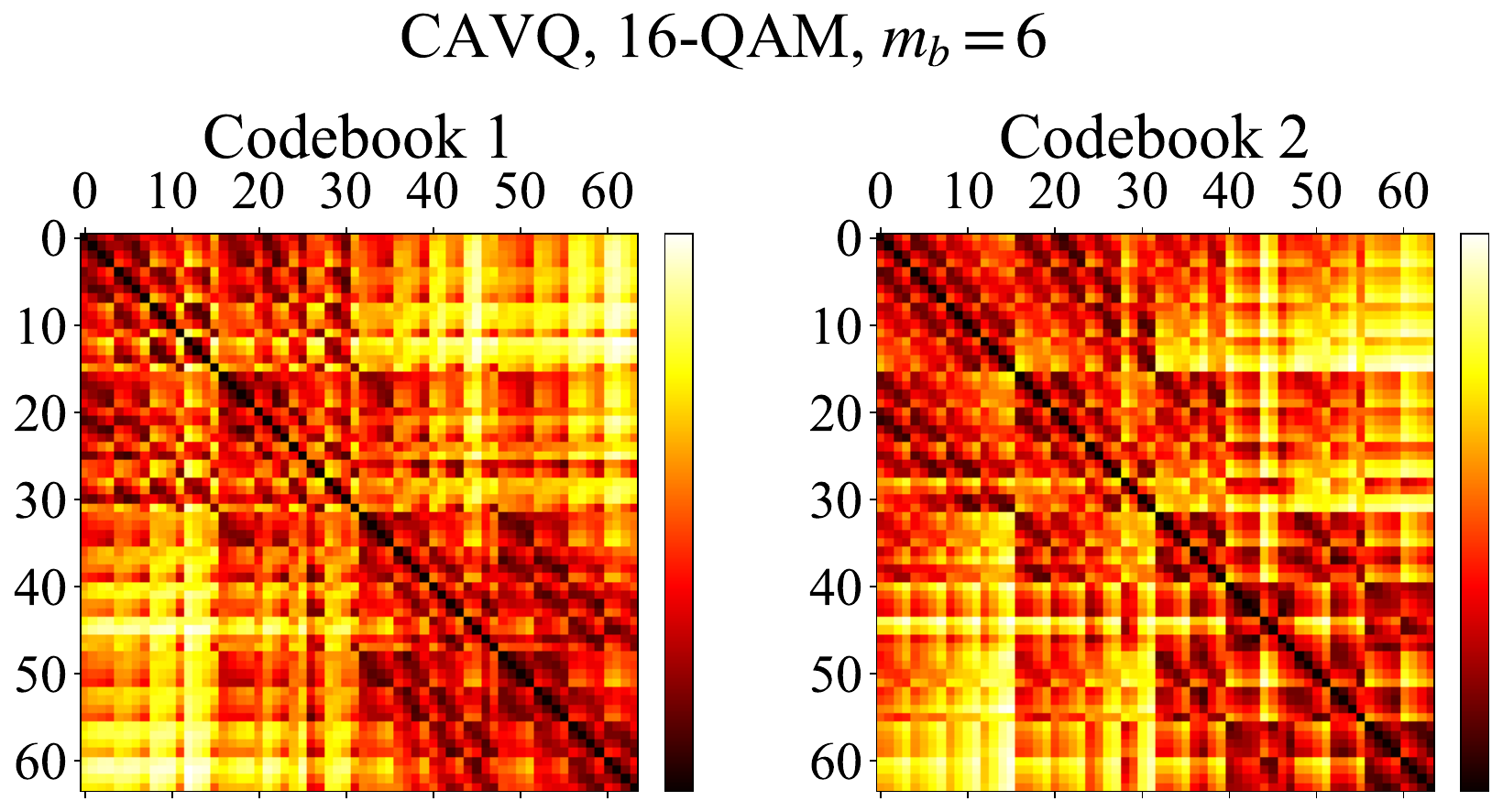}
  \end{minipage}
  
  \vspace{0.2cm}
  
  \begin{minipage}{0.45\textwidth}
    \includegraphics[width=\linewidth]{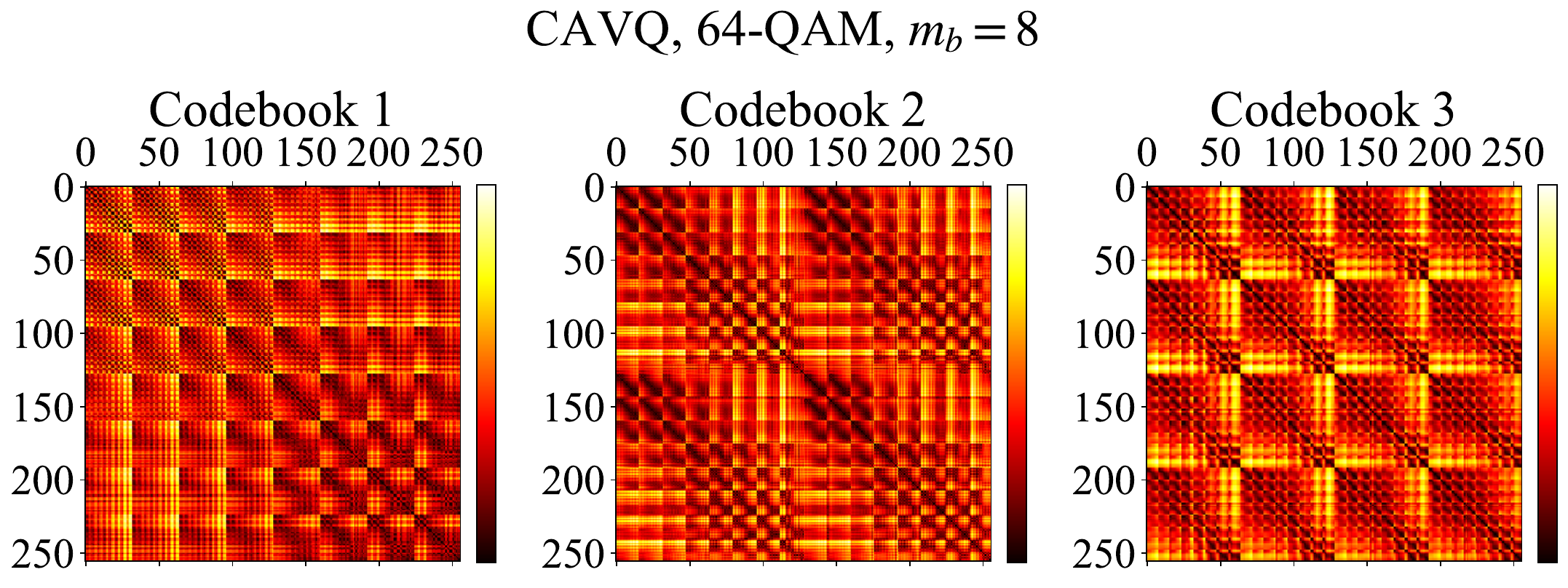}
  \end{minipage}

  \vspace{0.2cm}

  \begin{minipage}{0.45\textwidth}
    \includegraphics[width=\linewidth]{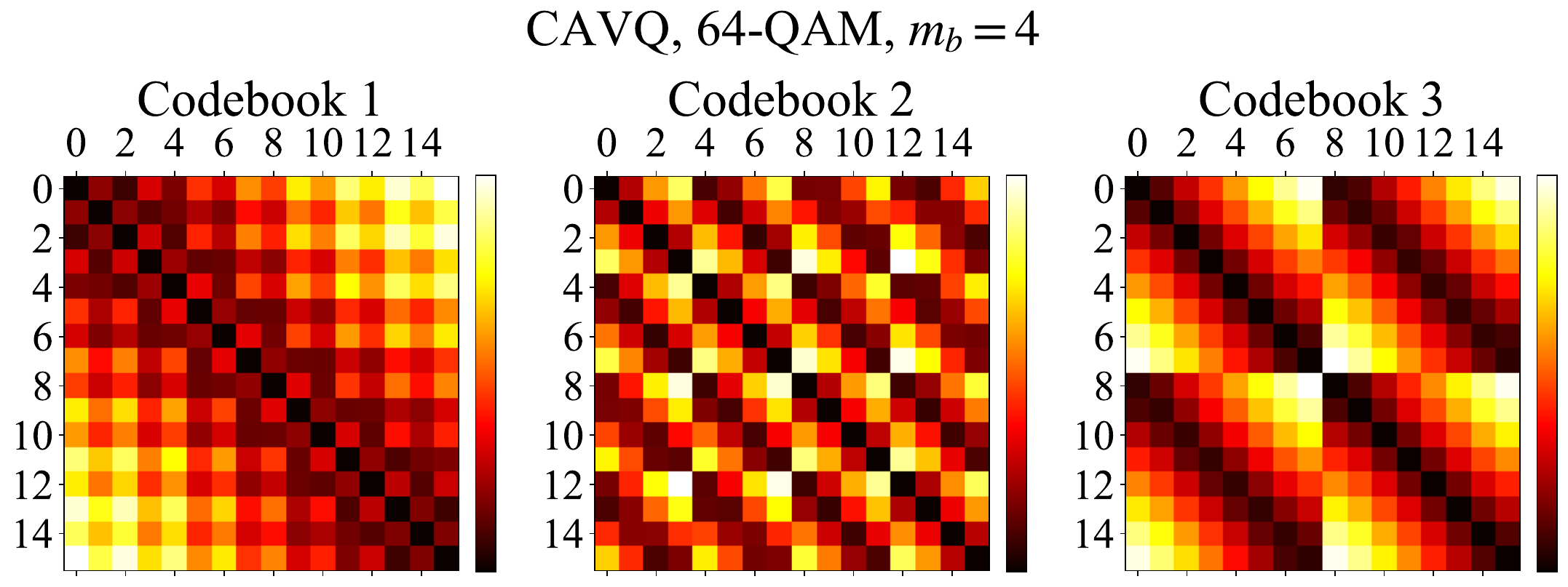}
  \end{minipage}
  
  \vspace{0.2cm}
  
  \begin{minipage}{0.33\textwidth}
    \includegraphics[width=\linewidth]{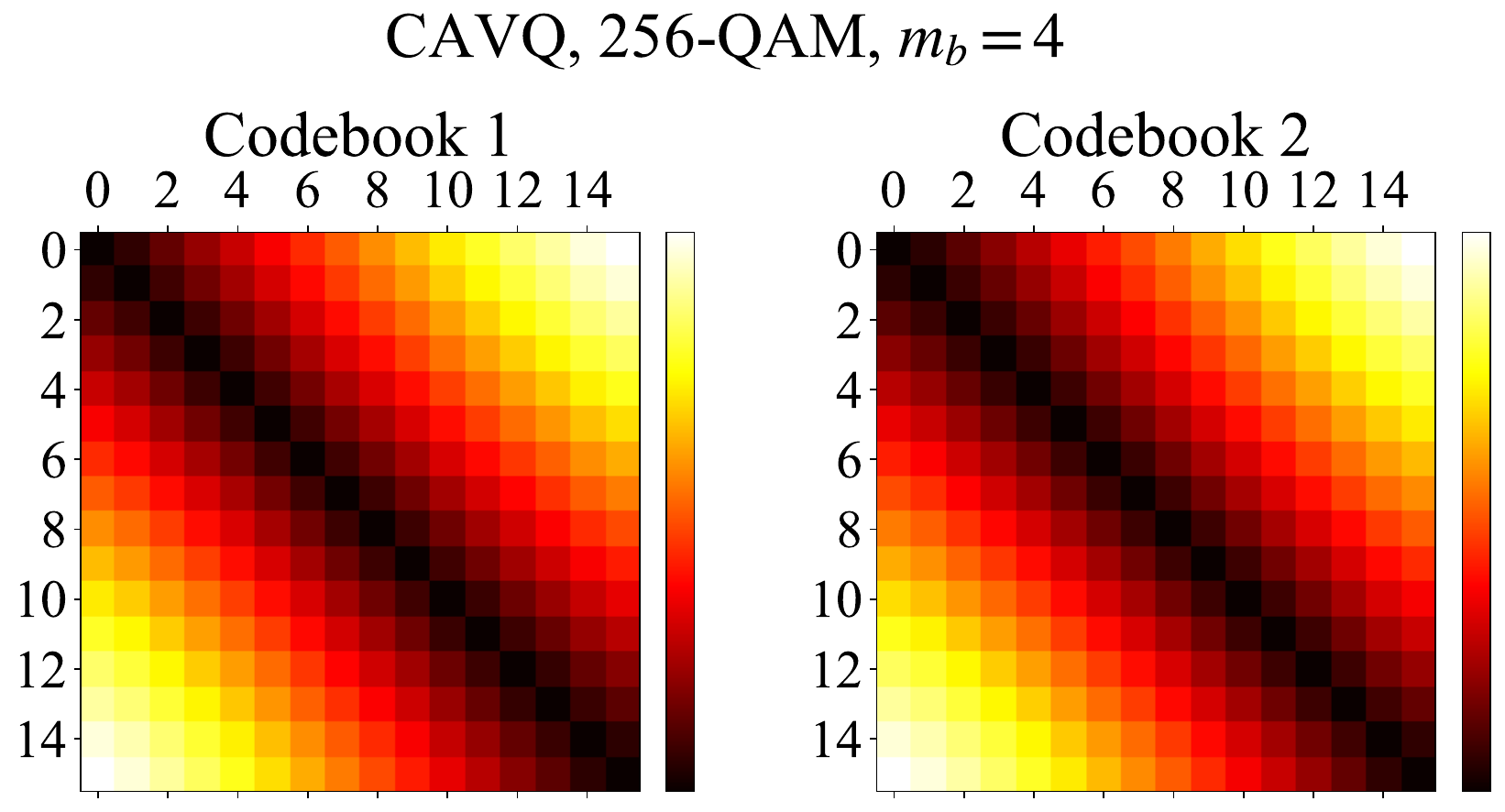}
  \end{minipage}
  \caption{Heatmaps of codeword distance matrices under mismatched codebook and modulation orders, generated by the multi-codebook algorithm with subsequent alignment via CAVQ. Each sub-codebook maintains structured patterns reflecting the estimated sub-channel transition probabilities.}
  \label{fig:multi_hotmap}
\end{figure}

In Fig.~\ref{fig:codebook_hotmap}, we compare distance heatmaps of modulation constellations and learned codebooks to illustrate the alignment achieved by our approach. For each modulation scheme, we first compute the pairwise Euclidean distances between constellation points to obtain a reference heatmap. Similarly, for a codebook with \(K\) \(d\)-dimensional vectors, we calculate the pairwise distances between codewords to form a matrix \(\mathbf{D} \in \mathbb{R}^{K \times K}\), where \(d_{ij} = \|\mathbf{m}_i - \mathbf{m}_j\|_2\). After min-max normalization, both constellation and codebook matrices are visualized as pseudo-color heatmaps, with warmer colors representing larger distances.

The comparison shows that vanilla VQ-VAE produces irregular and unstructured patterns, reflecting its inability to capture channel characteristics, whereas our CAVQ method yields constellation-like codebook structures that align closely with the modulation space. This alignment not only ensures that channel-induced errors between adjacent codewords cause only limited semantic distortion, but also effectively implements a form of soft quantization: instead of enforcing hard decision boundaries as in vanilla VQ-VAE, CAVQ leverages semantic proximity in the latent space and realizes quantization through statistical expectation, thereby bridging continuous and discrete representations and offering new design principles for future digital semantic communication systems.

Fig. \ref{fig:multi_hotmap} further illustrates the adaptability of our method when the codebook order does not match the modulation order. For example, with a codebook order of $m_b=4$ (16 codewords) under 64-QAM modulation, three independent sub-codebooks still preserve ordered structures. A similar consistency is observed even with 256-QAM, where two sub-codebooks maintain well-organized patterns. These results highlight the robustness and generalizability of CAVQ in establishing channel-aligned codebook distributions across diverse configurations.

\begin{figure}[t!]
  \centering
  \includegraphics[width=0.45\textwidth]{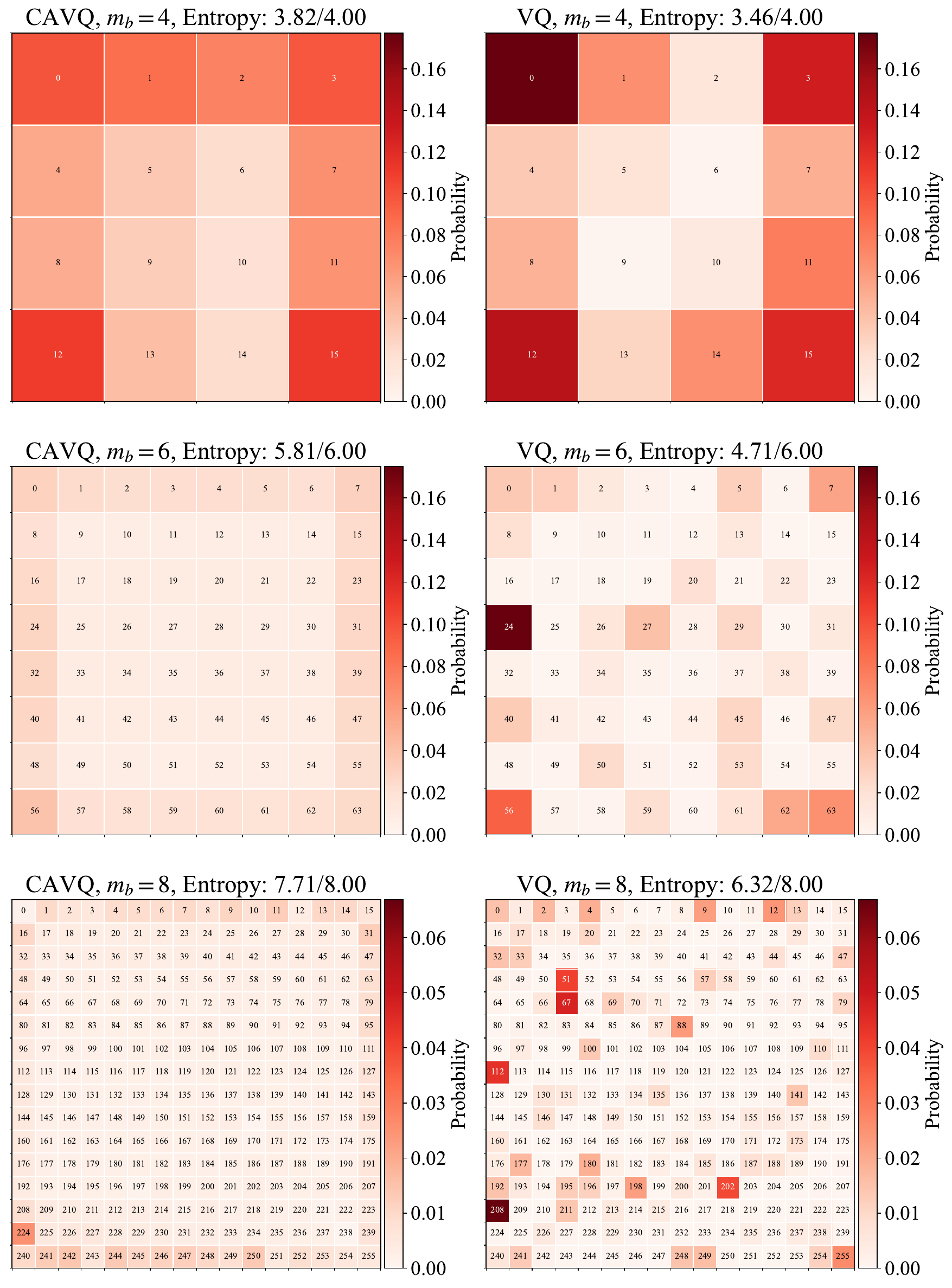}
  \caption{Codeword activation heatmaps under different codebook orders ($m_b=4,6,8$) corresponding to 16-QAM, 64-QAM, and 256-QAM. Each cell denotes the empirical usage frequency of a codeword, with darker colors indicating higher activation probability. Codeword positions are arranged according to the square QAM constellation.}
  \label{fig:freq_hotmap}
\end{figure}

Beyond the geometric alignment revealed by distance heatmaps, another critical perspective is the statistical distribution of codewords. Fig.~\ref{fig:freq_hotmap} presents codeword activation heatmaps under different codebook orders, corresponding to 16-QAM, 64-QAM, and 256-QAM. Each grid cell denotes the empirical activation frequency of a codeword over the test set, with darker colors indicating higher probabilities. The spatial arrangement of the cells mirrors the square QAM constellation, so the heatmaps can be directly interpreted as constellation-like activation maps. To ensure comparability, the colormaps are normalized row-wise, such that identical color depths correspond to the same activation probability across different methods, and the total frequency within each heatmap sums to one.

A clear contrast emerges between the two approaches. Vanilla VQ-VAE (right column) exhibits severe \emph{codebook collapse}: a few codewords dominate the representation space while most others are scarcely or never utilized. This imbalance arises because constellation-edge points have inherently lower error probabilities, making them more favorable for information encoding. However, such biased activation reduces effective entropy and weakens robustness. By contrast, our channel-aware method (left column) yields nearly uniform activation across all codewords, ensuring that the entire codebook contributes to representation.

These observations are quantitatively supported by the entropy values reported above each subfigure. Our method consistently achieves higher entropy (e.g., 7.71/8.00 at $m_b=8$) than vanilla VQ-VAE (6.32/8.00), confirming its ability to sustain richer semantic expressiveness while mitigating collapse. The combination of balanced utilization and elevated entropy highlights the importance of channel-aware design in preserving both capacity and robustness in digital semantic communication.

\subsection{PSNR Performance Comparison Across Benchmarks}

\begin{figure*}[htbp]
\centering
\includegraphics[width=\textwidth]{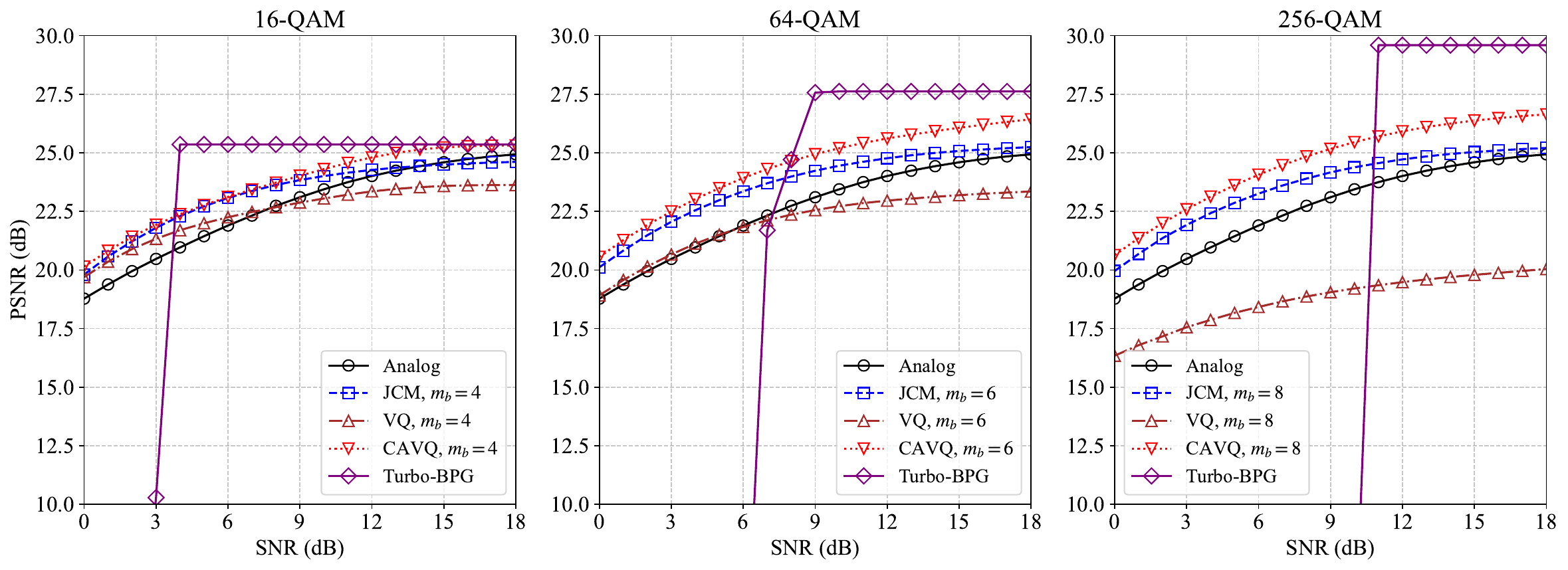}
\caption{Comparison of PSNR performance of different quantizers at various modulation orders ($l=3$ index depth).}
\label{fig:diff_quantizer}
\end{figure*}

\begin{figure*}[htbp]
\centering
\includegraphics[width=\textwidth]{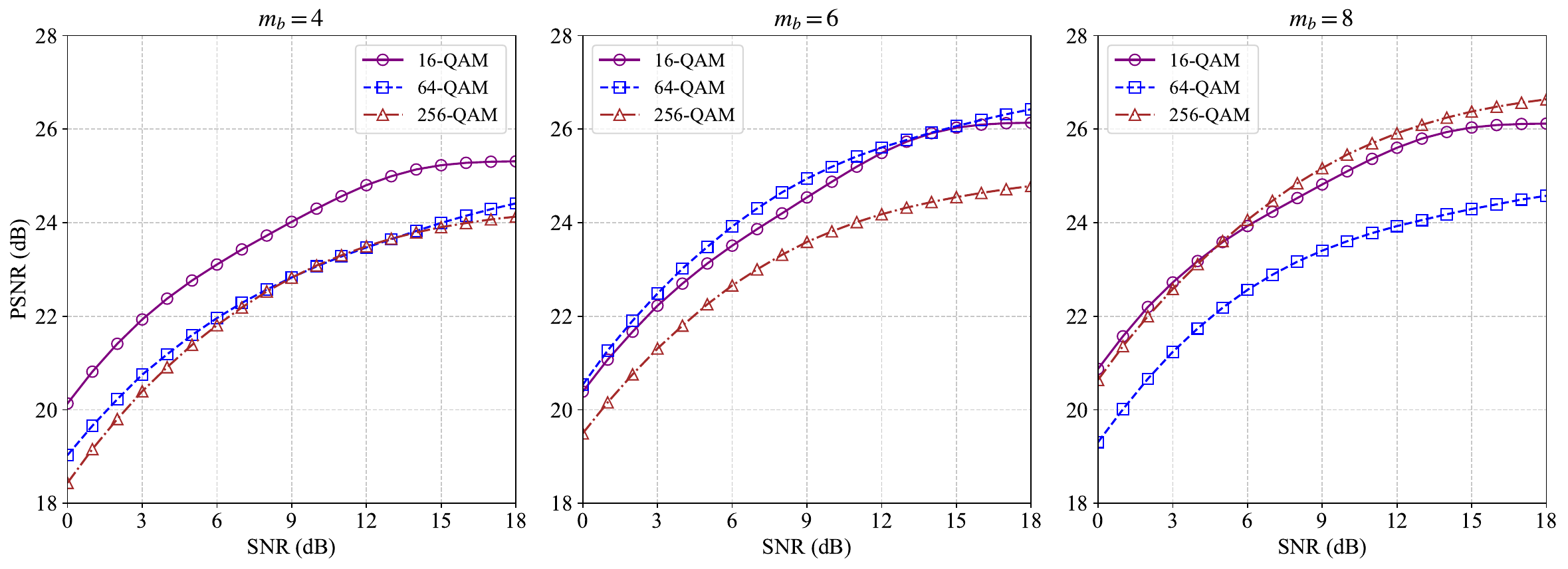}
\caption{Comparison of PSNR performance under different modulation orders $m_c$ and codebook orders $m_b$ ($l=3$ index depth).}
\label{fig:diff_const}
\end{figure*}

Figs.~\ref{fig:diff_quantizer} and \ref{fig:diff_const} jointly evaluate the impact of the codebook order $m_b$ and the modulation order $m_c$ on end-to-end semantic transmission performance under an index depth of $l=3$. These two parameters play complementary roles: $m_b$ governs the granularity of semantic quantization, determining how finely each feature is discretized, while $m_c$ controls the modulation efficiency at the physical layer. Their interaction directly shapes the tradeoff between semantic expressiveness and channel robustness.

Fig.~\ref{fig:diff_quantizer} compares four representative schemes—vanilla VQ-VAE (VQ), JCM, Turbo-BPG, and our proposed channel-aware VQ-VAE (CAVQ)—across different modulation orders. CAVQ consistently achieves the highest PSNR, and its advantage becomes increasingly prominent as the modulation order rises. This improvement originates from its ability to explicitly align the learned codebook distribution with the channel transition statistics, reducing the vulnerability of high-order constellations to symbol errors. In contrast, the vanilla VQ-VAE, which is unaware of channel effects, suffers substantial degradation at higher modulation orders. JCM achieves moderate robustness due to its joint source-channel training but remains limited by its semi-digital nature, while Turbo-BPG performs well at high SNR but exhibits a sharp cliff effect at low SNR due to its block-coding dependency.

It is also noteworthy that both CAVQ and JCM outperform the simple analog baseline. The reason lies in the evaluation setting: all methods share the same symbol rate, thus constraining the analog model to use the same latent dimensionality ($l=3$). This shallow latent space limits its ability to encode diverse semantic features. Conversely, quantized models benefit from structured codebook discretization, which enhances representational diversity while maintaining robustness against channel impairments. Consequently, even though the analog scheme avoids quantization noise, it falls behind in overall reconstruction fidelity.

\begin{figure}[htbp]
\centering
\includegraphics[width=0.48\textwidth]{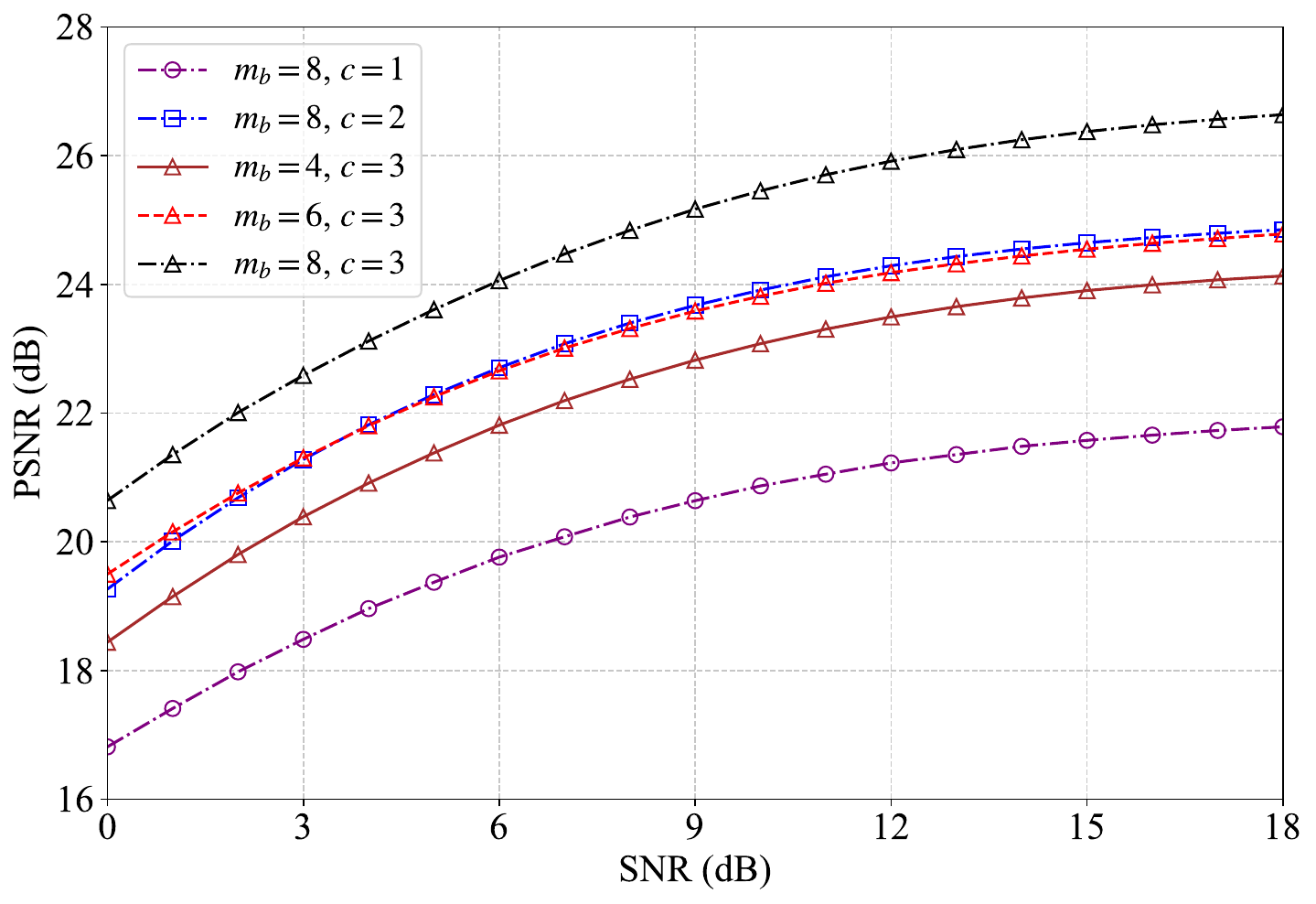}
\caption{PSNR performance under different codebook orders $m_b$ and index lengths $c$ with 256-QAM modulation. Both higher $m_b$ and longer $c$ improve performance, but mismatched configurations cause deviations from ideal scaling.}
\label{fig:diff_z}
\end{figure}

Fig.~\ref{fig:diff_const} further analyzes the interplay between $m_b$ and $m_c$ by fixing the overall transmission bit rate. For $m_b=4$, lower-order modulation (16-QAM) yields the best performance, since higher-order constellations suffer from excessive symbol errors despite better bandwidth efficiency. When $m_b=6$, 16-QAM and 64-QAM achieve comparable results, while 256-QAM remains suboptimal, though all cases outperform the $m_b=4$ baseline due to finer quantization. For $m_b=8$, 256-QAM overtakes other schemes in medium-to-high SNR regimes, indicating that the system benefits most when the codebook and modulation orders are well aligned. In contrast, combinations such as $(m_b=8,\text{64-QAM})$, $(m_b=6,\text{256-QAM})$, or $(m_b=4,\text{64-QAM})$ show weaker performance. This degradation arises because these configurations require splitting the index stream into multiple subsequences (as described in Algorithm~\ref{alg:remap}). Each subsequence is trained with a separate codebook prior estimated online via EMA, and the inherent stochastic bias in this estimation introduces distributional mismatch between the codebook and the channel transition probabilities.

Fig.~\ref{fig:diff_z} complements this analysis by exploring how the codebook order $m_b$ and index length $l$ jointly influence discrete semantic representation capacity under 256-QAM. Increasing either parameter consistently improves PSNR, as both enlarge the information space available per semantic feature. In essence, the overall representational capacity scales approximately with the product of $m_b$ and $l$, which together define the effective bit budget per feature. Configurations with similar total bit budgets, such as $(m_b=8, c=2)$ and $(m_b=6, c=3)$, achieve nearly identical PSNR, indicating that system expressiveness depends primarily on the total encoded information rather than on how it is distributed between quantization depth and sequence length. However, when the modulation and codebook orders are mismatched, as in the aforementioned multi-subsequence cases, performance deviates from this ideal scaling. The EMA-based prior estimation in the multi-codebook variant alleviates this misalignment to some extent but cannot fully eliminate the induced noise and bias, leading to suboptimal channel-codebook matching.

Two primary insights emerge from these results. Firstly, the roles of $m_b$ and $l$ in semantic representation are fundamentally alike, as they both regulate the total information capacity per feature. Secondly, stable training and robust performance are achieved when the codebook order and modulation order are aligned ($m_b \approx m_c$), which ensures a natural match between the learned codebook and the constellation structure. Consequently, in adaptive-rate digital semantic communication systems, rate control can be efficiently achieved by varying the index length $l$ while keeping $m_b$ fixed, simplifying both model design and optimization.

\section{Conclusion}\label{sec:omega}
This paper presented VQJSCC, a fully digital joint source–channel coding framework built upon the VQ-VAE architecture for discrete semantic communication. The proposed CAVQ incorporates channel transition statistics into the quantization process, while a multi-codebook mechanism enables channel-aware learning under mismatched codebook and modulation orders.
Experimental results show that VQJSCC consistently outperforms existing analog, semi-digital, and conventional JSCC methods across different SNRs and modulation schemes, effectively mitigating the digital cliff effect while preserving high reconstruction fidelity and efficiency. Further analysis reveals that system performance is primarily determined by the total information budget and achieves optimal robustness when the codebook and modulation orders are aligned, suggesting that rate adaptation can be efficiently realized by adjusting code length while maintaining codebook–modulation consistency.

{\appendices
\section{Proof of Theorem 1} \label{apx:t1}
To establish the error bound, we develop the squared norm through chained inequalities:

\begin{align}
\|\mathbf{x} - \hat{\mathbf{x}}\|_2^2 
&= \|(\mathbf{x} - g_\theta(\mathbf{Z})) + (g_\theta(\mathbf{Z}) - g_\theta(\hat{\mathbf{Z}}_q))\|_2^2 \nonumber \\
&\overset{(a)}{\leq} \left(\|\mathbf{x} - g_\theta(\mathbf{Z})\|_2 + \|g_\theta(\mathbf{Z}) - g_\theta(\hat{\mathbf{Z}}_q)\|_2\right)^2 \nonumber \\
&= \|\mathbf{x} - g_\theta(\mathbf{Z})\|_2^2 + \|g_\theta(\mathbf{Z}) - g_\theta(\hat{\mathbf{Z}}_q)\|_2^2 \nonumber \\
&\quad + 2\|\mathbf{x} - g_\theta(\mathbf{Z})\|_2 \cdot \|g_\theta(\mathbf{Z}) - g_\theta(\hat{\mathbf{Z}}_q)\|_2 \nonumber \\
&\overset{(b)}{\leq} 2\|\mathbf{x} - g_\theta(\mathbf{Z})\|_2^2 + 2\|g_\theta(\mathbf{Z}) - g_\theta(\hat{\mathbf{Z}}_q)\|_2^2 \nonumber \\
&\overset{(c)}{\leq} 2\|\mathbf{x} - g_\theta(\mathbf{Z})\|_2^2 + 2C_0^2 \|\mathbf{Z} - \hat{\mathbf{Z}}_q\|_2^2
\end{align}

Inequality (a) applies the triangle inequality to the vector sum. The doubling of coefficients in (b) arises from bounding the cross-term via $2ab \leq a^2 + b^2$. Final bound (c) utilizes the Lipschitz continuity assumption
\begin{align}
       g_\theta(\mathbf{Z}) - g_\theta(\hat{\mathbf{Z}}_q) \leq C_0\left( \mathbf{Z} - \hat{\mathbf{Z}}_q\right).
\end{align}
This assumption is realistically applicable because typical neural network elements such as fully connected layers, convolutional layers, and activation functions like ReLU or Sigmoid are demonstrably Lipschitz continuous. Thus, compositional architectures maintain Lipschitz continuity, ensuring that $C_0$ is finite for the decoder.

Finally, Theorem 1 is proved by applying the expectation
\begin{align}
    \mathbb{E}_{\mathbf{x,\hat{x}}}\left[ \|\mathbf{x} - \hat{\mathbf{x}}\|_2^2 \right] &\leq 2\mathbb{E}_{\mathbf{x}}\left[ \|\mathbf{x} - g_\theta(\mathbf{Z})\|_2^2 \right] \nonumber  \\
&\quad+ 2C_0^2 \mathbb{E}_{\mathbf{x},\hat{\mathbf{y}}}\left[ \|\mathbf{Z} - \hat{\mathbf{Z}}_q\|_2^2 \right].
\end{align}

\section{Proof of Proposition 1}\label{apx:p1}

\begin{align}
\mathcal{L}_{t}&= \mathbb{E}_{\mathbf{x},\hat{\mathbf{y}}}\left[ \|\mathbf{Z} - \hat{\mathbf{Z}}_q\|_2^2 \right] \nonumber\\
&\stackrel{(a)}{=} \mathbb{E}_{\mathbf{x}} \mathbb{E}_{\hat{\mathbf{y}}|\mathbf{x}} \left[ \|\mathbf{Z} - \hat{\mathbf{Z}}_q\|_2^2 \right] \nonumber\\
&\stackrel{(b)}{=} \mathbb{E}_{\mathbf{x}} \mathbb{E}_{\mathbf{y}|\mathbf{x}}\mathbb{E}_{\hat{\mathbf{y}}|\mathbf{x},\mathbf{y}} \left[ \|\mathbf{Z} - \hat{\mathbf{Z}}_q\|_2^2 \right] \nonumber\\
&\stackrel{(c)}{=} \mathbb{E}_{\mathbf{x}} \mathbb{E}_{\hat{\mathbf{y}}|\mathbf{y}} \left[ \|\mathbf{Z} - \hat{\mathbf{Z}}_q\|_2^2 \right] \nonumber\\
&\stackrel{(d)}{=} \mathbb{E}_{\mathbf{x}} \left[ \sum_{n=1}^N \mathbb{E}_{\hat{\mathbf{y}}|\mathbf{y}} \left[ \| \mathbf{z}_n - \mathbf{m}_{\hat{y}_{n}} \|_2^2 \right] \right] \nonumber\\ 
&\stackrel{(e)}{=} \mathbb{E}_{\mathbf{x}} \left[ \sum_{n=1}^N \mathbb{E}_{\hat{y}_n|y_{n}} \left[ \| \mathbf{z}_n - \mathbf{m}_{\hat{y}_{n}} \|_2^2 \right] \right] \nonumber\\ 
&\stackrel{(f)}{=} \mathbb{E}_{\mathbf{x}} \left[ \sum_{n=1}^N \sum_{k=1}^{K}P(\hat{y}_n\!=\!k\mid y_n)\| \mathbf{z}_n - \mathbf{m}_{k} \|_2^2 \right]
\end{align}

The above equalities (a)-(f) are elaborated as follows:
\begin{itemize}
    \item[(a)] Applies the law of total expectation to expand the joint expectation into nested conditional expectations over $\mathbf{x}$ and $\hat{\mathbf{y}}$.
    
    \item[(b)] Expands the expectation using the Markov chain $\mathbf{x} \rightarrow \mathbf{y} \rightarrow \hat{\mathbf{y}}$, where $\mathbf{y}$ denotes the codebook indices induced by $\mathbf{x}$.
    
    \item[(c)] Invokes the Markov property: (i) $\mathbf{y}$ becomes deterministic given $\mathbf{x}$, eliminating $\mathbb{E}_{\mathbf{y}|\mathbf{x}}$; (ii) Conditional independence $\hat{\mathbf{y}} \perp \mathbf{x}|\mathbf{y}$ reduces $\mathbb{E}_{\hat{\mathbf{y}}|\mathbf{x},\mathbf{y}}$ to $\mathbb{E}_{\hat{\mathbf{y}}|\mathbf{y}}$.
    
    \item[(d)] Decomposes the Frobenius norm into a summation over individual elements, then applies the linearity of expectation to interchange summation and expectation operators.
    
    \item[(e)] Utilizes the memoryless channel condition: Each $\hat{y}_n$ depends only on $y_n$, enabling per-element expectation decomposition.
    
    \item[(f)] Explicitly expands the inner expectation as a weighted sum over all possible quantization indices $k$, where weights correspond to codebook transition probabilities $P(\hat{y}_n=k|y_n)$.
\end{itemize}

}

\bibliographystyle{ieeetr}
\bibliography{ref.bib}

\end{document}